\documentstyle{article}
\newcommand{\be}{\begin{equation}}
\newcommand{\en}{\end{equation}}
\newcommand{\bea}{\begin{eqnarray}}
\newcommand{\ena}{\end{eqnarray}}  
\newcommand{\bean}{\begin{eqnarray*}}
\newcommand{\enan}{\end{eqnarray*}}
\newcommand{\bfig}{\begin{figure}}
\newcommand{\efig}{\end{figure}}
\newcommand{\Proof}{\smallskip\noindent{\bf Proof: }}
\newcommand{\qed}{\hfill $\Box$\smallskip}
\newcommand{\Remark}{\smallskip\noindent{\bf Remark: }}
\newtheorem{proposition}{Proposition}
\newtheorem{lemma}[proposition]{Lemma}

\newcommand{\csh}{{\rm csh }}
\advance\textheight by \topskip
\newdimen\figcenter
\newdimen\texpscorrection
\def\putfigurewithtex#1 #2 #3 #4
{
{
\figcenter=\hsize\relax
\advance\figcenter by -#4truecm
\divide\figcenter by 2
\vskip #3truecm
\noindent\hskip\figcenter
\includegraphics{#1}
{\hskip\texpscorrection\input #2 }
}
}     
\newlength{\psdim}
\newlength{\epsvoffset}
\newlength{\epshoffset}
\def\epsfig#1#2#3#4{
\epsvoffset= #4truecm  \advance \epsvoffset by #2\psdim
\divide \epsvoffset by 2
\epshoffset= \textwidth \advance\epshoffset by -#3\psdim
\divide \epshoffset by 2
\vskip \epsvoffset \hskip \epshoffset
\includegraphics{#1.eps}
}
\title{ A Lower Bound for Chaos on the Elliptical Stadium} 

\author
{Eduardo Canale\thanks{canale@fing.edu.uy \hfil
$^\star$roma@fing.edu.uy \hfil
$^\diamond$syok@mat.ufmg.br \hfil
$^\circ$sonia@mat.ufmg.br}
\hskip 0.7truecm
Roberto Markarian$^\star$
\\
Inst. de Matem\'atica y Estad\'\i stica 
``Prof. Ing. Rafael Laguardia'' \\
Fac. de Ingenier\'\i a, Universidad de la Republica\\
C.C. 30, Montevideo, Uruguay.
\and 
Sylvie Oliffson Kamphorst$^\diamond$
 \hskip 0.7truecm 
S\^onia Pinto de Carvalho$^\circ$
\\
Departamento de Matem\'atica, ICEx, UFMG\\
C.P. 702, 30161--970, Belo Horizonte, Brasil. 
}

\date{}

\begin{document}
\maketitle

\begin{abstract}
The elliptical stadium is a plane region bounded by a curve constructed by 
joining two half-ellipses,
with half axes $a>1$ and $b=1$, by two parallel segments of equal 
length $2h$.

V.~Donnay \cite{kn:don} proved that if $1<a<\sqrt 2$ and if $h$ is big 
enough than 
the corresponding billiard map has non-vanishing Lyapunov exponents almost
everywhere; moreover $h\to \infty$ as $a\to \sqrt 2$. In a previous paper 
\cite{kn:cmp} we found a bound for $h$ assuring the K-property for these 
billiards, for values of $a$ very close to $1$. 

In this work we study the stability of a particular family of periodic 
orbits obtaining a new bound for the chaotic zone for any value of 
$a<\sqrt 2$.

\end{abstract}

\section{Introduction}

The elliptical stadium is a plane region bounded by a curve $\Gamma$,
constructed
by joining two half-ellipses, with major axes $a>1$ and minor axes $b=1$,
by two straight segments of equal length $2h$ (see fig.\ref{fig:estadio}).

\bfig[h]
\putfigurewithtex figura1.ps figura1.ins 5 16 
\caption{The elliptical stadium.}
\label{fig:estadio}
\efig

The billiard on the elliptical stadium consists in the study of the free
motion of a point particle inside the stadium, being reflected elastically
at the impacts with $\Gamma$.
Since the motion is free inside $\Gamma$, it is determined
either by two consecutive points of reflection at $\Gamma$ or by the point of
reflection and the direction of motion immediately after each collision.

Let $s\in [0,L)$ be the arclength parameter for $\Gamma$ 
and the direction of motion be given
by the angle $\beta$ with the normal to the boundary at the impact point. 
The billiard defines a map $T$ from the
annulus ${\cal A}=[0,L)\times (-\pi/2,\pi/2)$ into itself. 
Let $(s_0,\beta_0)$ and $(s_1,\beta_1) \in {\cal A}$ be such that $T(s_0,\beta_0)=
(s_1,\beta_1)$ and that $\Gamma$ is $C^\infty$ in some neighborhood 
of $s_0$ and
$s_1$ (notice that $\Gamma$ is globally $C^1$ but not $C^2$ and piecewise
$C^\infty$).
Then, $T$ is a $C^\infty$-diffeomorphism in some neighborhoods of
$(s_0,\beta_0)$ and $(s_1,\beta_1)$. It also preserves the measure
$d\mu=\cos\beta \, d\beta \, ds$ (see, for instance, \cite{kn:mar}). 

$({\cal A},\mu,T)$ defines a discrete dynamical system, whose orbits are given by
$${\cal O}(s_0,\beta_0)=\{(s_n,\beta_n)=T^n(s_0,\beta_0), n\in Z \}\subset
A.$$  

So the elliptical stadium billiard defines, almost everywhere, a 
two-parameter 
family of
diffeomorphisms $T_{a,h}$ whose dynamics depend on the values of 
$a$ and $h$.
For instance, when $h=0$,  $T_{a,h}$ is integrable for every $a$
since we have the elliptical billiard.

In \cite{kn:don}, V.~Donnay proved that the elliptical billiard stadium
map $T_{a,h}$ has non-vanishing Lyapunov exponents almost everywhere if 
$1<a<\sqrt 2$ and $h$ is sufficiently large. He also proved that $h$ 
must go to infinity as $a$ approaches $\sqrt 2$. 
Donnay addressed a challenge:
``One could try to calculate bounds on these lengths.''

In \cite{kn:cmp}, we proved that if 
$1<a<\sqrt{4-2\sqrt 2}$, then $h>2a^2\sqrt{a^2-1}$ assures not only the 
positiveness of a Lyapunov exponent, but also ergodicity and the 
K-property. 
However, $2a^2\sqrt{a^2-1}$ does not seem to be an optimal lower bound 
for $h$. 
Numerical simulations exhibit chaotic phase spaces for values of 
$h$ smaller 
than this bound. Moreover, $\sqrt{4-2\sqrt 2}
\approx 1.082$ is far from $\sqrt 2$
and close to $1$, so in this case we are very close to the Bunimovich
stadium which is {\em chaotic} for all $h > 0$.

In this work, expanding and revising the work in
\cite{kn:can}, we study a family of periodic orbits, with 
any pair period $p \ge 4$, whose behavior looks generic. 
This means that in phase space, when they are elliptic, they
are surrounded by invariant curves which constitute {\em elliptic islands}
of positive measure which disappear as the orbits change 
from elliptic to hyperbolic. 
 
We will prove the existence of a curve $H=H(a)$, which diverges as
$a$ approaches $\sqrt 2$
such that above it all  
these periodic orbits are hyperbolic. 
Even though $H=H(a)$ may not be the optimal searched bound for chaoticity, 
it is at least a lower bound and 
seems not far from it. 
It also gives a very good answer to Donnay's challenge.

This paper continues in the following way: in section~2 we 
describe the family of periodic orbits. 
In section 3, we study their hyperbolicity. Section 4
consists on the definition of the bound $H(a)$. Section~5 contains
some concluding remarks.

\section{Pantographic Orbits}

On ${I\hspace{-.12truecm}R}^2$, we fix the origin on the center of the
elliptical stadium and take the $x$-axes containing the major half-axis of
the half-ellipses (see figure \ref{fig:estadio}).

Given $a$, $h$ and a positive integer $i$, an $(i,a,h)$-pantographic orbit,
denoted by $Pan(i,a,h)$, is a $(4+2i)$-periodic orbit, symmetric with 
respect to
the coordinate axes, alternating the impacts up and down, with exactly 4 
impacts at the half-ellipses (2 at each one, joined by a vertical path)
and $2i$ impacts at the straight parts ($i$ at each one) and crossing the 
$y$-axis only twice (see figure \ref{fig:pan}).

\bfig[h]
\putfigurewithtex figura2.ps figura2.ins 8.5 16 
\caption{Pantographic orbits of the elliptical stadium.}
\label{fig:pan}
\efig

The choice of those pantographic orbits was motivated by several remarks. 
The elliptical stadium billiard can be viewed as a perturbation of the 
elliptical billiard. 
The orbits of the elliptical billiard may be classified 
according to two different main features:
those that have an elliptical caustic 
and those with a hyperbolic caustic
(on the phase space of the
elliptical billiard, these  last orbits belong to 
invariant curves surrounding the elliptic period--2 point).
The invariant curves associated to the orbits with 
elliptical caustic are easily destroyed by the
perturbation (in the same way it happens in the Bunimovich stadium). 
This is
not so easy for those with hyperbolic caustic. 
So orbits that
have a chance to remain 
elliptic, after perturbation, must be 
close to
those trajectories with hyperbolic caustic.

On the other hand, it is known \cite{kn:mar}
that the elliptic character of an orbit can be given
by a relation between the curvature of the boundary at the impact points and
the total length 
of the trajectory. Since bounces on the straight parts of the 
elliptical stadium only change the length 
of the trajectory, the elliptic
behavior will depend fundamentally on the 
number of impacts with the
elliptical parts. But, if $a<\sqrt 2$, no trajectory with hyperbolic caustic,
on the elliptical billiard, can have more than two consecutive bounces on the
same half-ellipse \cite{kn:don}. 
Since while bouncing on the half-ellipse, any trajectory on
the elliptical stadium billiard behaves exactly as a trajectory on the
elliptical table, we must look for periodic orbits, close to orbits of the
elliptical billiard with hyperbolic caustic and bouncing twice at the
elliptical part of the stadium.

Between all the periodic orbits with this behavior, the 
pantographic have the following properties: \\
a) they exist for every even period $p\ge4$ and so can be studied as the 
period goes to infinity.\\
b) they can be explicitly localized and linearized.

Strengthening our choice of this 
family, in the numerical simulations we have carried out, 
those pantographic orbits appear as the last ones having {\em observable}  
KAM-like-islands.

\begin{proposition}
$Pan(i,a,h)$ exists for every $1<a<\sqrt{2}$ and $h>0$.
\end{proposition}

\Proof 
Let $P$ be the point of $Pan(i,a,h)$ located in the right half-ellipse
and on the first quadrant. Let $\lambda \in [0,\pi/2]$ be such that 
$P=(a\cos \lambda +h,\sin \lambda)$, and let $\beta > 0$ be 
the angle of the trajectory
by $P$, with the normal to the boundary. Using the obvious symmetries, 
it is easy to see that $P$ is a point of $Pan(i,a,h)$ if the straight line 
passing by $P$ with slope $\tan (\pi/2-2\beta)$ cuts the $y$-axis at
$(0,-i$). 
(see figure \ref{fig:equacao})

\bfig[h]
\putfigurewithtex figura3.ps figura3.ins 9 16
\caption{}
\label{fig:equacao}
\efig

So,
$$\tan 2\beta =\frac{h+a\cos \lambda}{i+\sin \lambda}.$$
And since $\tan \beta = \cos \lambda / (a\sin \lambda)$ we have
\be
a\tan \lambda-\frac{1}{a\tan \lambda}=
\frac{2(i\sqrt{1+{\tan}^2 \lambda}+\tan \lambda)}
{h\sqrt{1+{\tan}^2 \lambda} +a}
\label{eq:pan}
\en
It follows from the same arguments that a trajectory containing
the vertical piece from 
$ P'=  (a \cos \lambda + h, -\sin \lambda)$ to 
$ P = (a \cos \lambda + h, \sin \lambda)$
will cut the $y$-axis 
at $(0,-y)$, where
\be
y= y(t)= \frac{a^2 t^2-1}{2a t \sqrt{1+t^2}}(h\sqrt{1+t^2}+a)-
	\frac{t}{\sqrt{1+t^2}},
\label{eq:y}
\en
with $t=\tan \lambda$. 
Then finding a solution  of (\ref{eq:pan})
is equivalent to find a $t$ such that $y(t)=i$.

It is not difficult to verify that $\lim_{t\to 0} y(t) = -\infty$,
$\lim_{t\to \infty} y(t) = +\infty$ and 
$$
{dy \over dt} = {{\left( 1 + {a^2}\,{t^2} \right) \,
       \left( a + h\,(1+t^2)\,{\sqrt{1 + {t^2}}} \right)
       }\over {2\,a\,{t^2}\,{{\left( 1 + {t^2} \right) }^{{3\over 2}}}}}
\hbox{ $ > 0$ for all fixed $a>0$ and $h>0$ .}
$$
Then, as $y(t)$ is a continous
strictly increasing function running from
$-\infty$ to $+\infty$ as $t$ runs from $0$ to $\infty$, $y(t)=i$ has a
unique solution $t_i(a,h)$ for each integer $i$, for every
given $a>0$, $h>0$ and so (\ref{eq:pan}) has also a unique
solution $t_i(a,h) = \tan (\lambda_i(a,h))$.

On the other hand, as
we know \cite{kn:don} that for $a<\sqrt{2}$ no trajectory crossing the 
$x-axis$ between any two consecutive
hits with the boundary can have three consecutive impacts on the same
half-ellipse, we conclude that the next impact after $P'$ and $P$ on the
vertical piece of the trajectory as described above must be on the
straight part of the billiard.

Then if $s(\lambda_i)$ is the arclength corresponding to the point 
$P= (h + a \cos \lambda_i, \sin \lambda_i)$ of the stadium, and 
$0<\beta(\lambda_i)= \arctan {1/(a t_i)} <\pi/2$ the orbit of 
$(s(\lambda_i),\beta(\lambda_i))$ under the billiard map $T$ is 
$Pan(i,a,h)$.
So is the orbit of 
$(s(\lambda_i),\pi - \beta(\lambda_i))$.
\qed

{\bf Remark:} Using the same ideas,
the existence of those $Pan(i,a,h)$ can be  proved
for $1<a<2$, but we are only interested on $a<\sqrt 2$.

\begin{lemma}
Given $i$, $a$ and $h$, let $\lambda_i=\lambda_i(a,h)$ be the 
solution of (\ref{eq:pan}).
Then
$\lambda_i$ goes to $\arctan 1/a$ and $\beta(\lambda_i)$ goes to $\pi/4$ 
as $h$ goes to $\infty$;
$\lambda_i$ goes to $\pi/2$ as $h$ goes to $0$ 
for all $1<a<\sqrt{2}$.
\end{lemma}

\Proof 
The right side of equation (\ref{eq:pan}) goes to zero as $h \to \infty$ so
we must have in this limit, $a\tan \lambda = 1/(a \tan \lambda)$ and it 
follows that $\tan \lambda \to 1/a$ for all $i$.

To study the behavior when $h\to 0$ it is sufficient to study for $i=0$,
because it follows from the proof of Proposition~1 that if $i<j$ then 
$t_i<t_j$. For $i=0$, equation (\ref{eq:pan}) can be rewritten as:
\be
\frac{h}{a}(a^2-{\cot}^2 \lambda)\sqrt{1+{\cot}^2 \lambda}=[(2-a^2)+
{\cot}^2 \lambda]\cot \lambda
\label{eq:cot}
\en
and when $a<\sqrt 2$, $h\to 0, [(2-a^2)+{\cot}^2 \lambda]\cot \lambda\to 0$
and $t_i(0) \to \pi/2$. Thus, $t_i \to \pi/2$ as $h\to 0$.
\qed

\Remark
 Let us call {\it pantographic-like} orbits on the elliptical 
billiard the periodic trajectories that have vertical segments both at
left and right extremes. It would be amazing to compare the results in
Proposition 1 with the existence of those pantographic-like orbits.
As can be
seen in \cite{kn:cmp},
the $2n$-periodic pantographic-like 
orbit exists 
if $a>\overline a_n$ where $\overline a_n$ satisfies 
$\tan \pi/n ={2 \sqrt{\overline a_n^2-1}}/({\overline a_n^2-2})$. For
instance, there is no 4-periodic pantographic-like orbit if $a<\sqrt 2$, or
6-periodic if $a<2$.

\section{Hyperbolicity of the Pantographic Orbits}

\begin{proposition} \label{prop:hi}
For each $i$, let $\alpha_i = \sqrt{\frac{2+2i}{2+i}}$. 
\begin{enumerate}
\item For $i\geq 0$ if $\alpha_i<a<\sqrt 2$, there exists a unique $h_i(a)$
such that if $h<h_i(a)$, $Pan(i,a,h)$ is elliptic and if $h>h_i(a)$, 
$Pan(i,a,h)$ is hyperbolic.
\item For $i\geq 1$, if $1<a< \alpha_i$ , then $Pan(i,a,h)$ is hyperbolic for 
all $h>0$. 
\end{enumerate}
\end{proposition}

\Proof
For fixed $a < \sqrt 2$, $i$ and $h$, let $\lambda_i(a,h)$ be the solution
of (\ref{eq:pan}), $\beta$ be the angle, with the normal, of the outgoing
trajectory at $P=(a\cos \lambda_i+h,\sin \lambda_i)$ and $s$ the 
corresponding arclength.
So $T^{4+2i}\,(s,\beta)= (s,\beta)$ and to study the 
stability of this orbit we must analyze the eigenvalues of 
${\cal D}T^{4+2i} |_{(s,\beta)}$. 

Let
$(s_n,\beta_n)$ and $(s_{n+1},\beta_{n+1})$ be two consecutive impacts of
a trajectory with the two different half-ellipses
(with $k \ge 0$ impacts with the 
straight parts between them), or two consecutive impacts of
a trajectory with the same half-ellipse (with $k=0$),  then
(see, for instance,\cite{kn:mar}) 
$$
{\cal D} T^{k+1} |_{(s_n,\beta_n)} = 
{ (-1)^k \over \cos \beta_{n+1}}  
\left(
\begin{array}{cc}
l_{n,n+1} K_n - \cos \beta_n      &    l_{n,n+1} \\
K_n K_{n+1} ( l_{n,n+1} - \cos \beta_n - \cos \beta_{n+1}) & 
l_{n,n+1} K_{n+1} -
\cos \beta_{n+1} 
\end{array}\right)
$$
where $K$ stands for the curvature, and $l_{n,n+1}$ is the total length of
the trajectory between the two impacts with the half-ellipses. 

Then, using elementary geometry and the symmetries of the trajectory, 
we can write
${\cal D}T^{4+2i} |_{(s,\beta)}
= (M_1 M_2)^2 $ with
$$
M_j = { 1 \over \cos \beta}   
\left(
\begin{array}{cc}
l_j \, K - \cos \beta      &    l_j \\
K^2 \,  ( l_j - 2 \cos \beta)  & l_j \, K -
\cos \beta
\end{array}\right)
$$
and where 
$l_1 = 2 \sin \lambda_i$,
$l_2 = 2 \sqrt{(h+a\cos \lambda_i)^2+(i+\sin \lambda_i)^2}$ and 
$K =a/(a^2\sin^2 \lambda_i + \cos^2 \lambda_i)^{3/2}$.

Now if we define 
$$
\Delta_i(a,h) = \left( {l_1\, K \over \cos \beta} -1 \right ) \,
\left( {l_2 \, K \over \cos \beta} -1 \right ),
$$
we have that $Pan(i,a,h)$ is
{\em elliptic} if $0<\Delta_i(a,h)<1$,
{\em parabolic} if $\Delta_i(a,h)=0$ or $1$ and
{\em hyperbolic} if $\Delta_i(a,h)<0$ or $\Delta_i(a,h)>1$,
which means that the eigenvalues of ${\cal D}T^{4+2i} |_{(s,\beta)}$ are 
respectively purely imaginary and unitary, equal to 1, real and one 
bigger than 1 and the other 
smaller than 1 (remember that the system is conservative).

To study the function $\Delta_i(a,h)$, we need the following lemma:

\begin{lemma}
The function $\Delta_i(a,h)$ has the following properties for $h>0$ and
$1 < a < \sqrt{2}$:
\begin{enumerate}
\item $\Delta_i(a,h)>0$
\item $\frac{\partial \Delta_i}{\partial h}>0$ 
\item $\lim_{h\rightarrow +\infty}\Delta_i(a,h)=+\infty$
\item $\lim_{h\rightarrow 0}\Delta_i(a,h)=
L_i(a) = ( \frac{2}{a^2}-1)( \frac{2(i+1)}{a^2}-1)> 0$
\end{enumerate}
\label{lem:delta}
\end{lemma}

{\bf Proof of the lemma:}
If $a<\sqrt 2$ the half-osculating circles of the ellipse are entirely
contained inside the ellipse \cite{kn:cmp}, and so 
$l_1 > \frac{\cos\beta}{K}$. 
Since $l_2>l_1$,
$\Delta_i$ is the product of two positive factors and property~1 follows.

To prove property 2, we remark first that, from formula~\ref{eq:y}
we can derive implicitly $\partial t/ \partial h$ and show that it is negative.

We have 
$${l_1\, K \over \cos \beta} = {2 \over a^2 \sin^2 \beta + \cos^2 \beta}
= 2 {1 + t^2 \over 1 + a^2 t^2}
$$
and this is a decreasing function of $t$ if $a>1$.
So the first factor of $\Delta_i$ is a decreasing function of $t$.

For $0 < \lambda <\pi/2$ ($ t > 0$) $K$ is a decreasing function of 
$\lambda$, and so a decreasing function of $t$.
As $\tan \beta = 1/at$, $\cos \beta$ is an increasing function of $t$.
This implies that $K/ \cos \beta$ decreases with $t$.
Now 
$$ {\partial l_2^2 \over \partial \lambda} = 
{-a \sin \lambda \over 1 + \sin \lambda} \, (\tan 2 \beta - \tan \beta)
$$ 
is negative, as $\beta < \pi /4$.

Thus, $\Delta_i$ is a product of two decreasing functions of $t$, and so,
is an increasing function of $h$ which is property~2.
  
Property~3 is obvious since $l_2\to\infty$ as $h\to\infty$ and all the 
other quantities are bounded.

When 
$h\to 0$, $\lambda_i \to \pi/2, l_1\to 2, \cos\beta \to 1, K\to 1/a^2$ 
and $l_2\to 2(1+i)$,  which implies property~4.
\qed
 
Now we finish the proof of Proposition \ref{prop:hi}.
Given $i$, we know from the lemma above that for each $1< a < \sqrt{2} $,
$\Delta_i(a,h)$ is an increasing function of $h$ running from $L_i (a)$ to 
$\infty$. 
The function $L_i(a)$ decreases with 
$a$, $L_i(1)=1+2i>1$ if $i>0$ and $L_i(\sqrt 2)=0$ for every $i$. 
$L_i=1$ has the solution $a=\sqrt 2\sqrt{\frac{1+i}{2+i}} = \alpha_i$.

So, if $\alpha_i<a<\sqrt 2$, there exists a unique $h_i(a)$
such that $\Delta_i(a,h_i(a))=1$ and $h<h_i(a)$ implies 
$\Delta_i(a,h_i(a))<1$,
$h>h_i(a)$ implies $\Delta_i(a,h_i(a))>1$.

On the other hand, if $a < \alpha_i$, $\Delta_i(a,h) > 1$ for all $h$ and
the result follows. 
\qed

For each $i\geq 0$ fixed, and for all
$ \alpha_i < a < \sqrt 2$, formula (\ref{eq:pan})
and $\Delta_i(a,h) =1$ constitute a system equivalent to 
\begin{eqnarray}
&&
a^2\sin^3 \lambda + \frac{i}{2}(a^2 -1)\sin^2 \lambda
- \sin \lambda - \frac{i}{2}=0 \label{eq:poli}\\
&&
h_i(a) = {a \sqrt{1-\sin^2 \lambda} \over (a^2+1) \sin^2 \lambda - 1}
      \left ( 2 i \sin \lambda + 1 -  (a^2 - 1) \sin^2 \lambda \right )
\label{eq:sist}
\end{eqnarray}
where $\lambda = \lambda_i (a)$.

The values $\sin \lambda=1/a$ and $h=\sqrt{a^2-1}$ satisfy the equations 
above
for $i=0$. So, $h_0(a)=\sqrt{a^2-1}$ for 
$1<a<\sqrt 2$.

Now, let be $y = \sin \lambda_i(a)$. The problem of finding $h_i(a)$ is 
reduced to finding a root of the cubic polynomial
$$
P_{i,a}(y) = y^3 + {i\over 2 a^2} {(a^2 -1)} y^2 
- {1\over a^2} y - {i\over 2 a^2}
$$
in the interval $(0,1)$. $P_{i,a}(y) = 0$ can be rewritten as 
$$y \left ( y^2 - {1\over a^2} \right) = 
{i \over 2 a^2} \left ( 1 - (a^2 -1) y^2 \right)$$
The left hand side is a cubic polinomial with roots at $0, \pm 1/a$; it is 
positive for $y>1/a$ and negative in $(0,1/a)$. The right hand side
is a quadratic polynomial, with roots $\pm 1/\sqrt{a^2-1}$ and which is 
negative for $y>1/\sqrt{a^2-1}>1/a$. This implies that $P_{i,a}$ has only 
one positive real root and that this root belongs to
$(1/a,1/\sqrt{a^2-1})$. As the left hand side is 0 for $y=1/a$ and the
right hand side is positive, $P_{i,a}(1/a) <0$. 
However, 
$P_{i,a}(1) >0$ for $a>\alpha_i$ and $P_{i,a}(1) < 0$ for $a>\alpha_i$.
This implies that for each $i$, $P_{i,a}(y)$ has one and only one 
real root in $(0,1)$ for $\alpha_i < a < \sqrt 2$.

This root can be found by standard techniques: 
\begin{equation}
y_i(a)= 2 \, \sqrt{A} \,  
        \cos \theta - {i\, (a^2 -1) \over 6 \, a^2}
\label{eq:ycos}
\end{equation}
where 
\begin{eqnarray*}
A & = &
{  { 12\,{a^2} +  ({a^2} - 1)^2 \, {i^2} } \over { (6 a^2)^2 }}\\
B & = &
{ 2\,\left( {18\,{a^2} (1+ 2\,{a^2})\,i +   ({a^2}  -1 )^3 }\,{i^3} \right) 
\over (6 a^2)^3 } \\
\cos 3 \theta & = & {-B  / (2 A^{3/2})}
\end{eqnarray*}

Moreover, a more careful investigation shows that
$ {B / (2 A^{3/2})} < 1$. 
When $ {B  / (2 A^{3/2})} <  -1$, there is only one real root and
one should make use of the definition of cosine for  imaginary 
arguments so ``$\cos$''
is changed into 
``$\csh$''. For $-1 <  {B  / (2 A^{3/2})} \le 1$, there are 3 real roots and
we choose $ 0 \le \theta \le \pi/3$,
in order to have $y_i(a) > 0$.

One can also write (\ref{eq:ycos}) as
$$
y_i (a) = 
  \,  {A\over {C^{1\over 3}}}
 + \, C^{1\over 3}  - 
 {{\left( -1 + {a^2} \right) \,i}\over {6\,{a^2}}}
$$
where $C= (\sqrt{B^2 - 4 A^3}- B)/2 $, and when it is imaginary, the 
choice of the logarithm branch is the same as the choice of $\theta$ above.

If we introduce this value of $\sin \lambda_i(a)$ in (\ref{eq:sist}), we 
obtain explicit formulae for the $h_i(a)$.
These functions are plotted in
figure~\ref{fig:hii}. 

\bfig[h]
\putfigurewithtex figura4.ps figura4.ins 11.5 16
\caption{Graphs of $h_i$}
\label{fig:hii}
\efig

\section{A lower bound for the chaotic zone}

For $i=0$, $h_0(a) = \sqrt{a^2 - 1}$ is a strictly increasing function in 
$(1,\sqrt{2})$. Figure~\ref{fig:hii} shows that this is also 
true for $i=1,~2$.
For $i\geq 3$ we have the following:

\begin{proposition} \label{prop:dhda}
For each fixed $i\ge3$,
 $\frac{d h_i}{d a} > 0$.
\end{proposition}

\Proof 
For $i> 0 $, we have
$$  
h_i(a) =  
{{2\, a\,y\,{\sqrt{1 - {y^2}}}
     \over {-1 + \left( 1 + {a^2} \right) \,{y^2}}}
\,\left( i + {{-1 + {a^2}\,{y^2}}\over i} \right)}
$$  
where $y = y_i(a)$ is the only root of $P_{i,a}(y)$
in $(1/a,1)$. Then, we have that
$$
{d y \over d a}=
  {{- a\,{y^2}\,\left( i + 2\,y \right) }\over 
    {  i\,y \,( {a^2} - 1 ) + 3\,{a^2}\,{y^2} - 1}} < 0.
$$ 

Now
$$
{dh_i \over da} 
= 
{\partial h_i \over \partial a} + {\partial h_i \over \partial y}{dy \over da}
$$
can be written  as
$$
{ dh_i \over da} 
={\partial h_i \over \partial a} 
\left( 1 + {a \over y (1-y^2)} \, {d y \over d a} \right)
+
\left( {\partial h_i \over \partial y} - 
{a \over y (1-y^2)}\, {\partial h_i \over \partial a}
\right) \, {dy \over da}.
$$
As
\bean
{\partial h_i \over \partial a} &=&
 {{2\, y\,{\sqrt{1 - {y^2}}}\,
\left( 1  - {y^2} - 2\,{a^2}\,{y^2} + 3\,{a^2}\,{y^4} + {a^4}\,{y^4} - \, 
        {i^2}\, ( 1 - {y^2} + {a^2}\,{y^2}  ) \,
         \right) }
\over {i\,{{\left( -1 + {y^2} + {a^2}\,{y^2} \right) }^2}}}\\
{\partial h_i \over \partial y} &=&
{{2\, a\,\left( 
1  - {y^2} - 2\,{a^2}\,{y^2} + 3\,{a^2}\,{y^4} + {a^4}\,{y^4} 
+ 2 \,{a^2}\,{y^4}- 2\,{a^2}\,{y^6} - 2\,{a^4}\,{y^6} 
- {i^2} ( 1 - {y^2} +  {a^2}\,{y^2} )
\right) }\over 
    {i\,{\sqrt{1 - {y^2}}}\,{{\left( -1 + {y^2} + {a^2}\,{y^2} \right) }^2}}}
\enan
we obtain that
$$
{\partial h_i \over \partial y} - 
{a \over y (1-y^2)}\, {\partial h_i \over \partial a}=
{{- 4\,{a^3}\,{y^4}}\over 
    {{i \, \sqrt{1 - {y^2}}}\, \left( - 1 +  \,{y^2} + {a^2}\,{y^2} \right) }}
< 0 
$$ 
and
$$
1 + {a \over y (1-y^2)} \, {d y \over d a} = 
{ 
 1 + {y^2}\, (  3\,{a^2}\,{y^2} - 1 - {a^2} )
+ 
i \, y \, (1  - {y^2} + {a^2}\,{y^2} ) 
\over 
- \left( 1 - {y^2} \right) \,\left( 
 i\,y\, ( {a^2} - 1)   +  3\,{a^2}\,{y^2} -1    
\right)
} < 0 
$$

Now we will show that $\partial h_i / \partial a < 0 $ for $i \ge 3$, so 
$dh_i / da > 0$ for $i\ge 3$.

$$
{\partial h_i \over \partial a} =
 {{2\, y\,{\sqrt{1 - {y^2}}}
\over {i\,{{\left( -1 + {y^2} + {a^2}\,{y^2} \right) }^2}}}
\left( D - {i^2}\, E
\right) }
$$
where
\bean
D &=& (1  - {y^2} ) + {a^2} \,{y^2} \, ( 3{y^2} -1)  + 
          {a^2}\,{y^2}\, ({a^2}\,{y^2} -1)  > 0\\
E &=&   1 + ( {a^2} - 1) \,{y^2}  > 0
\enan

Now 
$$
 E \ge 1 + {y^2 \over 3} \ge 1 + {1\over 6} = {7\over 6} > 0
$$
as
$ \sqrt{2} \ge a \ge \alpha_1 = 2/\sqrt{3}$, for $i\ge 1$,  and
$ 1 \ge y \ge 1/a \ge 1/\sqrt{2}$.

On the other hand
$$
0  < D \le {1\over 2} + 2 a^2 \, y^2 + a^2 \, y^2 = {1\over 2} + 3 a^2 \, y^2 
\le {1\over 2} + 6 = {13 \over 2} \ .
$$

So if
$$
i \ge  3 >  \sqrt{ 39 \over 7} \ge \sqrt{D \over E}
$$
$\partial h_i / \partial a < 0 $ and the result follows.\qed

We were also able to found the asymptotical behavior of the $h_i$ at 
$a=\sqrt{2}$.

\begin{proposition}
If $a=\sqrt{2}$ then 
$\displaystyle
\lim_{i\to \infty} {h_i^2(\sqrt{2}) \over 4 i} =1
$.
\label{prop:lim}
\end{proposition}

\Proof
For $a=\sqrt{2}$ we have 
$$
P_{i,\sqrt{2}}(y) = y^3 + {i\over 4} y^2 - {1\over 2} y - {i\over 4}
$$
so its only positive real root goes to 1 as $i$ goes to $\infty$. 
It follows that
$$
\lim_{i\to \infty} \sin \lambda_i(\sqrt{2}) = 1 \ .
$$
Since $P_{i,\sqrt{2}}(\sin \lambda) = 0$,
$$
\sin \lambda \left ( \sin^2 \lambda - {1\over2} \right) =
{i \over 4 } \, \cos^2 \lambda \ . 
$$
The left hand side goes to 1/2 as $i$ goes to $\infty$ implying
$$
\lim_{i\to \infty} {i\over 2} \, \cos^2 \lambda_i(\sqrt{2}) = 1 \ .
$$
As
$$
h_i^2(\sqrt 2) = 
{2 \cos^2 \lambda \over 3 \sin^2 \lambda - 1}
\left( 2 i \sin \lambda + \cos^2 \lambda \right )^2
$$
using the limits above, gives the desired result. \qed

Proposition \ref{prop:lim} tells us 
that there exists an $N$ such that if $i>N$, 
$h_i^2(\sqrt 2)/4 \approx i$. Plotting 
$h_i^2(\sqrt 2)/4 \times i$,  
we have observed the same linear
behavior also for small values of $i$.

For each $1<a<\sqrt 2$ there exists a $j$ such that $\alpha_{j-1}\leq
a<\alpha_j$. So, $Pan(i,a,h)$ is hyperbolic for $i\geq j$ and is elliptic if
$0\leq i<j$ and $h<h_i(a)$. There exists, then, only a finite number of
$h_i$'s defined for this value of $a$ and we can define the announced lower
bound by $H(a)=\max_{i<j}~\{h_i(a)\}$. Proposition \ref{prop:lim} implies that 
$H(a)\to \infty$ as $a\to\sqrt2$. 
$H(a)$ can be seen on figure~\ref{fig:hii}.

\section{Final remarks}

Clearly above the curve $H(a)$ all pantographic orbits are
hyperbolic and bellow it some have eigenvalues in the unit complex circle.
Generically, KAM theory establishes the existence of positive measure 
elliptic islands surrounding those orbits and our numerical simulations 
corroborate this result. In this paper we have not {\em proved}
the existence of such islands.
 
On the other hand, above this bound, although we can prove that all 
the orbits of this family are hyperbolic,
we can not assure the non existence of other elliptic periodic orbits that
could be surrounded by positive measure sets of invariant curves and
having vanishing Lyapunov exponents. As a matter of fact, we have observed
other periodic orbits of $T_{a,h}$ but the elliptic islands around them seem
to disappear for values of $h<H(a)$.

A numerical case study which seems generic for this problem is presented in 
figures \ref{fig:1.24a}, \ref{fig:1.24b} and \ref{fig:1.24c}.
There, we show the phase space associated for $a=1.24$ 
fixed and different values of $h$. For this value of $a$ we have that 
$\alpha_2=\sqrt{1.5}<1.24<\sqrt{1.6}=\alpha_3$ (see fig.~\ref{fig:hii}) so 
$Pan(i,1.24,h)$ are hyperbolic for any $h>0$ and $i\ge 3$.
The pantographic orbits of period 4, 6 and 8 ($Pan(0,1.24,h)$, 
$Pan(1,1.24,h)$ and $Pan(2,1.24,h)$) are the only relevant pantographic orbits 
as they are elliptic for small $h$. In fact we have $h_0 \approx 0.7332$, 
$h_1 \approx 1.0236$, $h_2 \approx 0.6770$, 
$h_2<h_0<h_1=H(1.24)$. 

Note that for very small $h$ (in this example, 0.1), a 
very rich structure of elliptic islands (the white holes on 
fig.~\ref{fig:1.24a})
can be observed; some of these islands correspond to other periodic orbits.

\bfig[h]
\epsfig{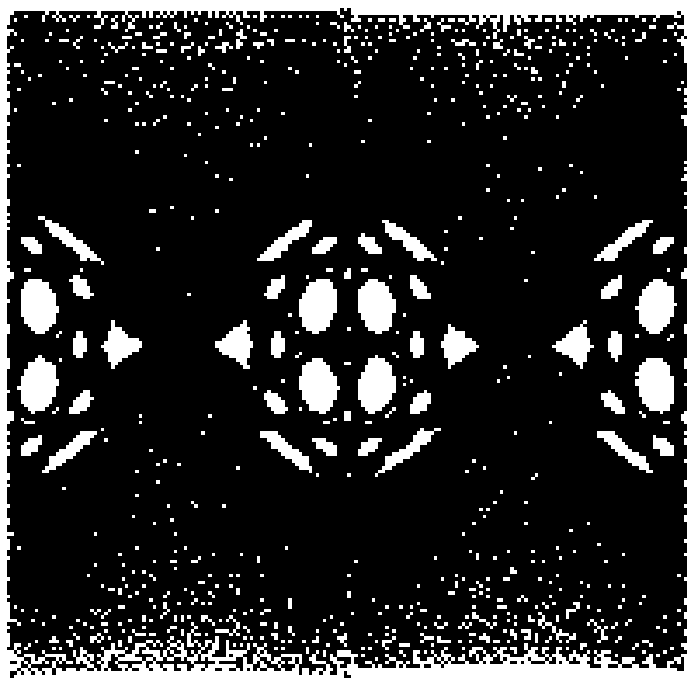}{193}{198}{7.5}
\caption{$a=1.24$, $h=0.1$. 150,000 iterations of a single 
initial condition.}
\label{fig:1.24a}
\efig

As $h$ is increased (fig.~\ref{fig:1.24b}),
they gradually disappear and only the pantographic 
islands seem to remain at $h=0.45$. Then, at $h=0.73>h_2$, the eight
islands around $Pan(2,1.24,h)$ have disappeared and at  
$h_0<h=0.75<h_1$ only the six islands around $Pan(1,1.24,h)$
can be seen, since $Pan(0,1.24,h)$ has also become hyperbolic. 
In each case, 
the region outside the elliptic islands seems to be a single
ergodic component as it is filled up by a single orbit.

\bfig
\epsfig{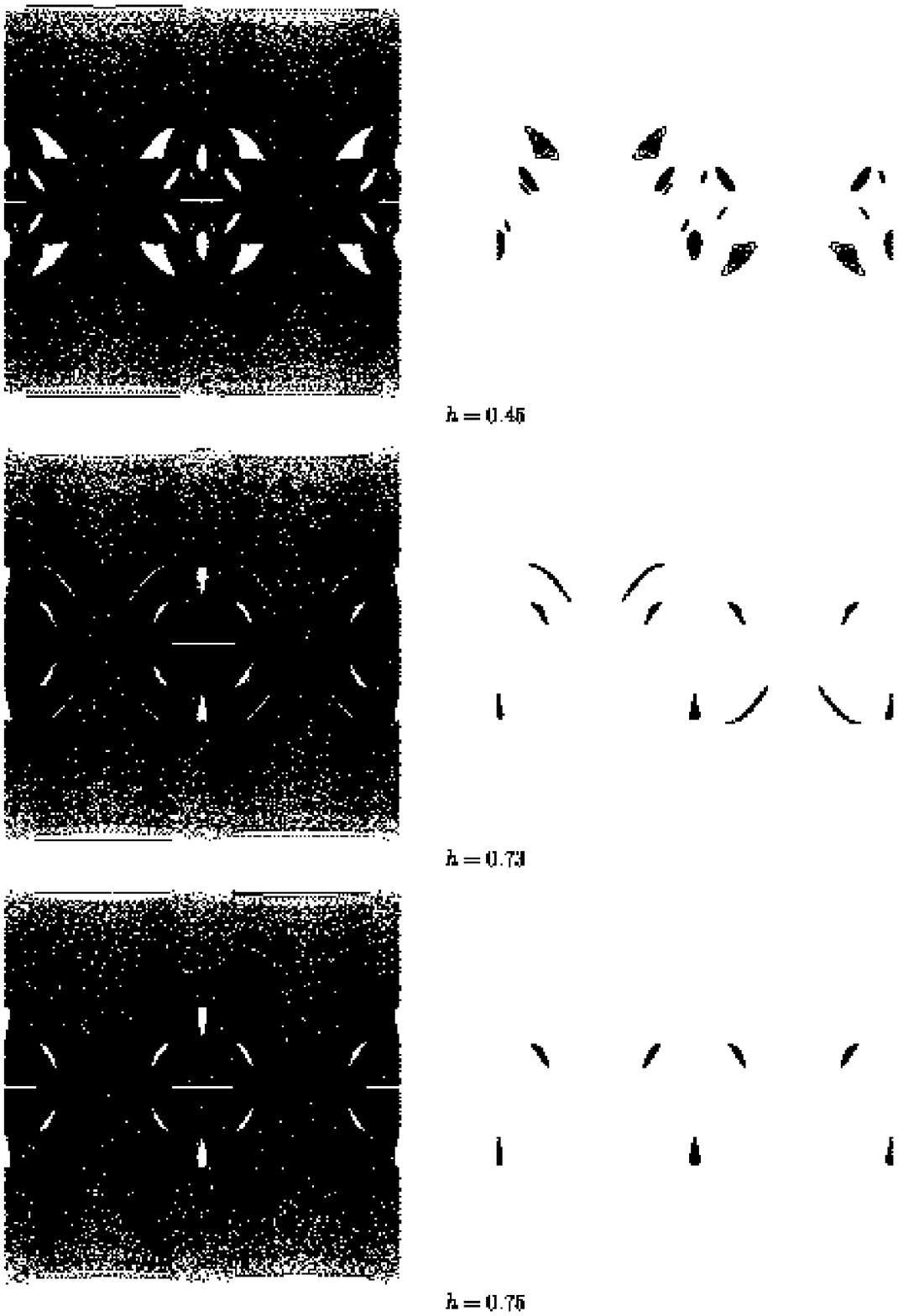}{653}{452}{20}
\caption{$a=1.24$ and different $h$'s. On the left side,
150,000 iterations
of a single initial condition and. On the right, the iteration of 
a few initial conditions close
to the elliptic pantographic orbits.
}
\label{fig:1.24b}
\efig

For $h=1.05>h_1=H(1.24)$ (fig.~\ref{fig:1.24c})
the system seems to be ergodic. 

\bfig
\epsfig{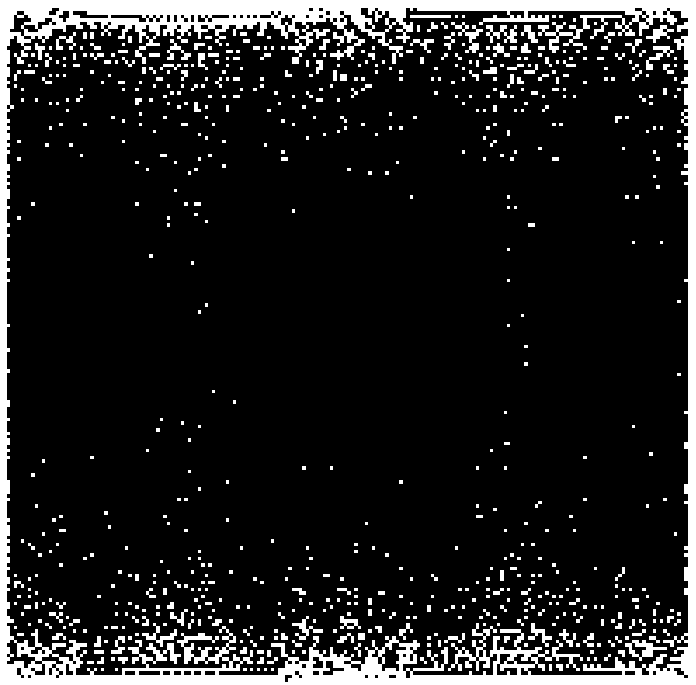}{195}{201}{6.5}
\caption{$a=1.24$, $h=1.05$. 150,000 iterations of a single 
initial condition.}
\label{fig:1.24c}
\efig

As we can see on figure~\ref{fig:hii}, the order of extinction of the
pantographic elliptic islands depends on the value of $a$. As shown above, for
$a=1.24$, the order is $i=2$, then $i=0$ and then $i=1$. But, for values of
$a$ a little bigger than 1.24, it can be 0,2,1 or even 0,1 and 2.

 However one can learn from 
figure~\ref{fig:hii}, 
that as $a$ approaches $\sqrt{2}$ the last islands to disappear
correspond to orbits with long period.
This is expressed in proposition~\ref{prop:lim}.
So in this system, and this fact seems also to occur in other problems,
the obstruction to ergodicity seems to be existence of elliptic islands
around  orbits with arbitrarily large period 
and so the existence of an arbitrarily large number
of very small islands. 
Those islands, even though summing up into a positive measure region,
can be invisible in a numerical simulation.

We would like to point out that the pantographic orbits also seem to be
the most important ones from the point of view of their focusing properties.
In \cite{kn:cmp}, the existence of a positive Lyapunov exponent for the
elliptical billiard was obtained through the study of the behavior of the 
caustic pencil, or the tangent vector to the invariant curves of the 
elliptical 
billiard. In the pantographic orbits, after hitting a half-ellipse twice, in 
the vertical portion of the trajectory, it crosses to the other half-ellipse. 
At this moment the focusing distance of the caustic pencil can be very large.
This lack of focalization should be compensated by a larger 
traveling distance, 
so a bigger $h$, in order to have a splitting of neighboring trajectories. 
However, as $a$ approaches $\sqrt{2}$ this focusing distance may tend to 
$\infty$. The loss of ellipticity by all the pantographic orbits caused by 
increasing $h$ indicates that the behavior of the caustic pencil may be 
controlled at this point and one should be able to prove the existence of 
positive Lyapunov exponents.

\vskip 1.2truecm
{\bf Acknowledgements.} We thank Cooperaci\'on Regional Francesa, FAPEMIG 
(Brasil), Programa de Recursos Humanos del PEDECIBA/CONICYT and 
Cooperaci\'on Internacional of the Universidad de la Rep\'ublica (Uruguay) 
for 
sponsoring visits of RM and SPC. SOK thanks the Department of Mathematics at
Boston University, where part of this work was done under a CAPES (Brasil) 
grant. EC and RM were partially supported by CSIC, Univ. de la Rep\'ublica 
(Uruguay); SOK and SPC by CNPq (Brasil).


\begin{thebibliography}{99}

\bibitem{kn:can}
E.~Canale, R.~Markarian: 
Simulaci\'on de billares planos. Anales IEEE, 
Segundo Seminario de Inform\'atica en el Uruguay, 71-96 (1991)\\
E.~Canale:
Informe Final del Proyecto de Iniciaci\'on en la Investigaci\'on 
(Comissi\'on Sectorial de Investigaci\'on Cient\'\i fica)
{\em Simulaci\'on de Sistemas Din\'amicos} (1995)
\bibitem{kn:don}
V.~J.~Donnay:
Using integrability to produce chaos: billiards with positive
entropy. Comm. Math. Phys. {\bf 141}, 225-257 (1991)
\bibitem{kn:mar}
R.~Markarian: Introduction to the ergodic theory of plane billiards. 
In: {\em Dynamical Systems}, Santiago de Chile, 1990. 
Bam\'on, Labarca, Lewowicz, Palis, eds. Harlow: Longman, 327-439 (1993)
\bibitem{kn:cmp}
R.~Markarian, S.~Oliffson Kamphorst, S.~Pinto de Carvalho:
Chaotic Properties
of the Elliptical Stadium. Comm. Math. Phys. {\bf 174}, 661-679 (1996)  
\end{thebibliography}
\end{document}